\documentclass[preprintnumbers,amsmath,amssymb]{revtex4}
\usepackage{amsmath}    
\usepackage{graphicx}   
\usepackage{color}

\pdfoutput=1
\begin{document}

\title{Solving the hard problem of Bertrand's paradox}

\author{Diederik Aerts}
\affiliation{Center Leo Apostel for Interdisciplinary Studies and Department of Mathematics, \\ Brussels Free University, Brussels, Belgium}
\email{diraerts@vub.ac.be}

\author{Massimiliano Sassoli de Bianchi}
\affiliation{Laboratorio di Autoricerca di Base, Lugano, Switzerland}
\email{autoricerca@gmail.com}

\date{\today}

\begin{abstract}
Bertrand's paradox is a famous problem of probability theory, pointing to a possible inconsistency in Laplace's principle of insufficient reason. In this article we show that Bertrand's paradox contains two different problems: an ``easy'' problem and a ``hard'' problem. The easy problem can be solved by formulating Bertrand's question in sufficiently precise terms, so allowing for a non ambiguous modelization of the entity subjected to the randomization. We then show that once the easy problem is settled, also the hard  problem becomes solvable, provided Laplace's principle of insufficient reason is applied not to the outcomes of the experiment, but to the different possible ``ways of selecting'' an interaction between the entity under investigation and that producing the randomization. This consists in evaluating a huge average over all possible ``ways of selecting'' an interaction, which we call a \emph{universal average}. Following a strategy similar to that used in the definition of the Wiener measure, we calculate such universal average and therefore solve the hard problem of Bertrand's paradox. The link between Bertrand's problem of probability theory and the measurement problem of quantum mechanics is also briefly discussed.  
\end{abstract}


\maketitle

\section{Stating the problem}

The so-called (by Poincar\'e) \emph{Bertrand's paradox} describes a situation where a same probability question receives different, apparently correct, but mutually incompatible, answers. To date, more than a century after Joseph Bertrand enounced in 1889 his famous riddle,~\cite{Bertrand} it remains a controversial issue in the foundations of probability theory (see Refs.~\onlinecite{Jaynes, Marinoff, Porto, Rowbottom, Shackel} and the references cited therein), calling into question the validity of \emph{Laplace's principle of insufficient reason}~\cite{Laplace} (saying that possibilities of which we have equal ignorance should be assigned equal probabilities).  

It is the purpose of the present article to clarify the content of this longstanding issue,  showing that Bertrand's problem cannot undermine Laplace's principle of insufficient reason (also called \emph{principle of indifference}, by Keynes~\cite{Keynes}), provided that the former is posed in non ambiguous terms and that the latter is applied at the level where the randomization (i.e., the actualization of potentials) truly occurs. When this is done, Bertrand's problem becomes solvable, in the sense that it admits a unique solution.

As it will be explained in the last section of the article, our solution to Bertrand's problem stems from an operational and realistic approach to the foundations of physical theories, known as the \emph{Geneva-Brussels} approach, which originated in the work of J. M. Jauch~\cite{Jauch1968} and C. Piron,~\cite{Piron1976} in Geneva, and was further developed by one of us, and his collaborators, in Brussels.~\cite{Aerts1982, Aerts1986, Aerts1998, Aerts1999a, Aerts1999b, AertsDurt1994, Aertsetal1997a, Aertsetal1999b} More specifically, the conceptual and mathematical analysis at the basis of our result, emerged from a recent investigation of the notion of \emph{universal measurement}, characterizing the most general possible condition of lack of knowledge in a measurement situation.~\cite{AertsSassoliOne, AertsSassoli,AertsSassoli-Bloch} This because when facing the problem of finding a physically transparent and mathematical precise definition for the average subtended by such universal measurements, we had to openly address the difficulty expressed by Bertrand's paradox, so that by finding a solution to the former we could also identify a way to tackle and solve the latter. 

We start by recalling the original formulation of the problem, using Bertrand's own words:~\cite{Bertrand}\\

\noindent {\bf Bertrand's Question (BQ)}: We draw \emph{at random} a chord onto a circle. What is the probability that it is longer than the side of the inscribed equilateral triangle?
\\

\noindent A specific physical realization of the above question is for instance the following:\\

\noindent {\bf Experimental Realization of BQ (EQ)}: What is the probability that a stick, \emph{tossed by a human being} onto a unit circle drawn on the floor, will define on it, when it intersects it, a chord of length $L$ greater than $\sqrt{3}$?
\\

\noindent Of course, the above EQ is to be understood in an idealized sense. The stick is assumed to be long enough, so that when tossed onto the circle the probability that it will intersect it in a single point, without being tangent to the circle, is negligible. Also, it is assumed to be sufficiently thin, so as to define points on the circle with sufficient resolution, and of course  the circle is assumed to be drawn on the floor with sufficient precision, etc. 

An important difference between the above two questions is that in the BQ the term ``at random'' is used, whereas in the EQ this term does not appear. This is because the act of \emph{drawing} is a process which is perfectly under the control of a designer. If for instance a designer would decide in advance what chord s/he will draw, say a chord passing from the center of the circle, s/he can do so without difficulty, so that the probability  that its length will be greater than the side of the inscribed equilateral triangle is equal to one. But this wasn't the kind of situations Bertrand wanted to investigate: he wanted the chord to be drawn by a person who would not have chosen it in advance, and expressed this condition by saying that the chord is drawn \emph{at random}, i.e., in a non predeterminable way. 

Consider now the EQ, which doesn't mention the term ``at random.'' It says that the stick has to be \emph{tossed by a  human being}, but it doesn't say that it has to be tossed \emph{at random}. This is because, if it would say so, it would not constitute an experimental realization of BQ. Indeed, it is precisely the purpose of the experimental realization to translate the term ``at random'' into a specific process which would be representative, or sufficiently representative, of what a random process is. This is done in the  EQ by asking a human being to toss the stick onto a circle drawn on the floor. 

Contrary to the act of drawing a chord, a human being tossing a stick on the floor doesn't have the ability to predetermine what will be the obtained chord. This because there are a number of fluctuating factors determining the final location and orientation of the stick which are totally beyond the \emph{control power} of the human who throws the stick. And of course, it is implicitly assumed here that the person performing the action does not have extraordinary abilities, and does not try to take control of the tossing process, by whatever means, so as to increase the frequency of certain outcomes.

\section{Bertrand's ``solutions''} 

Let us  recall the three different answers to BQ that were  historically proposed by Bertrand. These provide three different values for the sought probability -- which we denote ${\cal P}(L>\sqrt{3})$ -- and since they all appear to be correct, at least at first sight, this is why Poincar\'e named BQ, and similar questions, a paradox. 

\begin{figure}[!ht]
\centering
\includegraphics[scale =0.65]{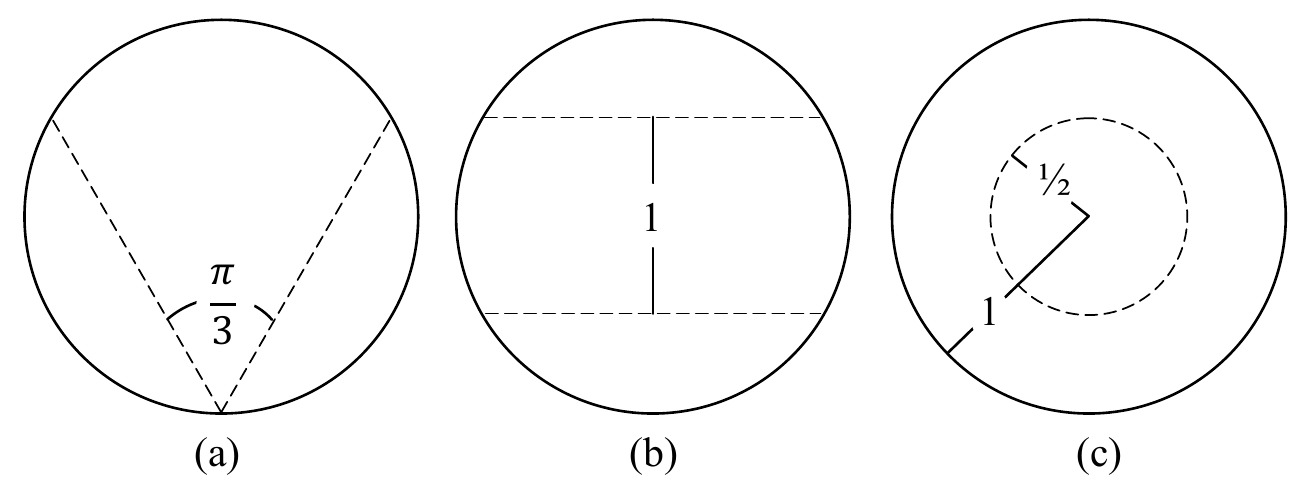}
\caption{A schematic description of the three different solutions proposed by Bertrand to its famous problem, giving the probabilities: (a) ${\cal P}(L>\sqrt{3})={1\over 3}$; (b) ${\cal P}(L>\sqrt{3})={1\over 2}$; (c) ${\cal P}(L>\sqrt{3})={1\over 4}$.
\label{circle}}
\end{figure}

(a) The first proposed solution consists in choosing an arbitrary point on the circle, considering it as one of the vertexes of an inscribed equilateral triangle. This point, describing one of the two points of intersection of the chord with the circle, is kept fixed, whereas the second point is varied (so that the chord moves like a pendulum). One then observes that by considering all possible points on the circle, the chord will rotate of a total angle of $\pi$, but that only the chords lying within the arc subtended by an angle of ${\pi\over 3}$ at the vertex satisfy Bertrand's condition $L>\sqrt{3}$; see Fig.~\ref{circle} (a). Thus, one finds: 
\begin{equation}
{\cal P}(L>\sqrt{3})={{\pi\over 3}\over \pi}={1\over 3}.
\end{equation}

(b) The second proposed solution consists in first choosing an arbitrary direction, and then considering chords which are all parallel to that direction. Then, moving the chords along the circle, one observes that those intersecting its diameter in its central segment of length $1$, satisfy Bertrand's condition $L>\sqrt{3}$; see Fig.~\ref{circle} (b). Being the total length of the diameter equal to $2$, one finds: 
\begin{equation}
{\cal P}(L>\sqrt{3})={1\over 2}.
\end{equation}

(c) The third proposed solution consists in  choosing an arbitrary point inside the circle, considering it as the middle point of the chord. Then, moving this point within the entire area of the circle, one  observes that all the chords having their middle point within the internal circle of radius ${1\over 2}$, i.e., within the circle inscribed in the equilateral triangle, satisfy Bertrand's condition $L>\sqrt{3}$; see Fig.~\ref{circle} (c). Being the area of such circle ${\pi \over 4}$, whereas the area of a circle of unit radius is $\pi$, one finds: 
\begin{equation}
{\cal P}(L>\sqrt{3})={{\pi\over 4}\over \pi}={1\over 4}.
\end{equation}

Consider now these three solutions in the perspective of the above EQ. We can observe that when a human tosses a stick onto a circle drawn on the floor, the process is not in general reducible to one of the processes described in the three solutions proposed by Bertrand. Take solution (a): it is clear that when one tosses a stick, one usually does not produce a movement which first defines a point on the circle, and then a rotation of the stick around this point. By this we are not saying  that such movement would be impossible to realize, but certainly it would be very difficult to voluntarily obtain. Similarly, take solution (b): when tossing a stick a person does not generally produce a trajectory which first defines an orientation for the stick, and then, once such orientation is obtained, moves the stick along the circle without altering its acquired orientation. Again, we are not saying that such movement would be impossible to obtain, but certainly almost impossible to produce at will. The same remains true, of course, for solution (c): a human operator does not usually have the ability to throw a stick onto a circle in such a way that the middle point of the chord it produces also corresponds to the middle point of the stick.

What we are here emphasizing, in accordance with the recent analysis of Rowbottom,~\cite{Rowbottom} is that none of the three solutions proposed by Bertrand correspond to a valuable experimental realization of his theoretical question. Indeed, they all correspond to ``ways of tossing the stick'' which are much too specific in comparison to the ``fully uncontrolled way of tossing a stick'' described in the EQ. Let us call these three different ``ways of tossing a stick'' onto a circle $\rho_1$, $\rho_2$ and $\rho_3$, respectively. Then, the three solutions proposed by Bertrand should be more conveniently written as three different conditional statements: ${\cal P}(L>\sqrt{3}|\rho_1)={1\over 3}$, ${\cal P}(L>\sqrt{3}|\rho_2)={1\over 2}$, and ${\cal P}(L>\sqrt{3}|\rho_3)={1\over 4}$, where ${\cal P}(L>\sqrt{3}|\rho_i)$ denotes the \emph{conditional} probability that the chord length is greater than $\sqrt{3}$, knowing that the two points on the circle, defining the chord, are selected according to the ``$\rho_i$-way of tossing the stick,'' $i=1,2,3$.

Now, it is perfectly clear that different conditions will produce, in general, different probabilities. Therefore, when we write the solutions proposed by Bertrand as conditional probabilities, the paradox immediately disappears.

\section{Ways of tossing}

Let us use the above first element of clarification to go deeper into BQ. For this, we need to consider more attentively what happens exactly when a human being tosses a stick onto a circle drawn on the floor. We can observe that as soon as the stick has left the hand of the thrower, its evolution becomes fully deterministic, in the sense that by knowing the position and orientation of the stick, its translational and rotational velocities, in the precise moment it loses contact with the hand (and assuming of course we also know the intrinsic properties of the stick, like its mass, density, length, volume, etc.), we would be able to predict,  in principle, what the final location of the stick on the floor will be, and therefore  the length of the corresponding chord onto the circle.    

In other terms, as soon as the stick has left the hand of the thrower, we have solid reasons to believe that there is a unique possible chord, so that if we would apply, in that moment, the Laplacean principle of insufficient reason~\cite{Laplace} (saying that the probability for an event is the ratio of the number of cases favorable to it, to the number of all cases possible, when nothing leads one to expect that any one of these cases should occur more than any other), we would obtain only two possible values for the probability, $0$ or $1$, as is clear that when the stick is in the air the number of favorable cases is either $0$ or $1$, and the number of all cases is just $1$. Thus, the chord does not get selected during the flight of the stick, but before the stick gets separated from the hand of the thrower. Strictly speaking, the thrower doesn't select a chord, but an interaction between her/his hand and the stick, which then produces, in a perfectly deterministic way, the final chord onto the circle. 

At this point, it is useful to recall what Edwin Jaynes emphasized in his famous 1973 paper:~\cite{Jaynes} that when we solve a problem we should not use any information that is not given in the statement of the problem. If we apply this reasonable principle of ``maximum lack of knowledge'' to our issue, we can observe that the protocol described in the EQ does not specify in which way the stick has to be tossed. This ``no specification'' specification is precisely what is implied by the term ``at random'' in BQ:  no level of control should be exerted by the agent, when s/he throws the stick, so as to produce a specific ``way of tossing the stick,'' instead of another. 

By a ``way of tossing a stick'' we mean  here a process which, if it would be repeated at each throw, would produce a  specific statistics of outcomes for the chords, characteristic of that specific way of tossing them. And, as we mentioned above, the three solutions proposed by Bertrand are precisely an expression of three different ``ways of tossing a stick,'' able to produce three different probabilities for the outcomes. 

When we face a situation of maximum lack of knowledge, this is precisely the moment we should apply the venerable principle of insufficient reason (or of indifference). Usually, the principle is applied at the level of the outcomes, and in many problems this is certainly sufficient to obtain the correct answer. However, logically speaking, the principle should be more consistently applied at the level where the different possibilities are actualized. In our case, this is the level where the human operator selects a specific \emph{way of tossing the stick}, which then produces the (still non-deterministic) selection of a given interaction between her/his hand and the stick, and ultimately gives (deterministically)  rise to a chord onto the circle. 

Since neither the BQ, nor its experimental realization EQ, lead us to expect that any one of these ``ways of tossing a stick'' should occur more frequently than any other, we have to consider them as all equally possible, and see if we can exploit this condition of maximum lack of knowledge to obtain a unique answer to the EQ, and therefore to the BQ.

\section{The ``easy'' and ``hard'' problem}

To realize the above program, we need to mathematically modelize the different ``ways of tossing a stick,'' and then consider the problem of calculating a uniform average over them. It is worth observing that this modelization will be very sensitive to the exact way the BQ is stated. Here we face a source of possible ambiguity which is not related to the fundamental problem of understanding what the term ``at random'' means, but to the more circumstantial problem of making sufficiently precise what is the specific nature of the entity which is subjected to the random process. 

In his question, Bertrand mentions a chord, i.e., a \emph{straight-line abstract entity}, which is the mathematical modelization of material entities like sticks, straws, needles, etc. But Bertrand could have written his question in the following, apparently similar, but certainly not equivalent way:\\

\noindent {\bf  Modified Bertrand's Question (BQ2)}: We draw \emph{at random} two points onto a circle (i.e., onto the boundary of a disk). What is the probability that the chord passing through these two points is longer than the side of the inscribed equilateral triangle?
\\

\noindent Clearly, this modified version of Bertrand's question does not admit a physical realization in terms of sticks. Indeed, the question now indicates that two points, and not a straight line, are the entities subjected to the randomization. Therefore, a convenient physical realization of the above modified question is for instance the following:\\

\noindent {\bf Experimental Realization of BQ2 (EQ2)}: What is the probability that two pebbles, \emph{tossed by a human being} onto a unit circle drawn on the floor, will define on it, when they both intersect the circle's curve, a chord of length $L$ greater than $\sqrt{3}$?
\\

\noindent It is important to realize that BQ and BQ2 are non equivalent questions, as they refer to different entities -- straight lines, in the original formulation, and pairs of points in our proposed variant -- admitting very different physical realizations. Therefore, it is to be expected that these two non equivalent questions will receive different answers. And of course, if we uncritically mix non equivalent questions (erroneously considering them as equivalent) into a single fuzzy question, by doing so we will artificially create a paradox.

In other terms, it is important to distinguish the ``modelization problem'' from the ``randomization problem.'' The solution of the modelization problem only requires a sufficiently precise definition of the entity which is subjected to the randomization. The solution of the randomization problem, on the other hand, requires to understand what a situation of \emph{maximum lack of knowledge} really implies, in relation to an entity which has been previously unambiguously defined. 

By inspection of Bertrand's solution (a), we can see that this confusion was already present in the original statement of the problem. Indeed, solution (a) consists in a procedure where one selects, in sequence, two points on the circle, to which only subsequently one associates a chord. This is what one would naturally do by tossing two pebbles, and, as previously mentioned, this is a process which would be extremely difficult (if not impossible) to realize by tossing a single stick. 

On the other hand, Bertrand's solution (b) describes a two-step procedure where an orientation is first chosen for the chord, and then an orthogonal displacement for it is produced. This is what one would naturally do (although usually not in sequence) when tossing a stick, but is extremely difficult (if not impossible) to realize by tossing two uncorrelated pebbles. 

So, to proceed in our analysis of Bertrand's question, we first need to decide what is the nature of the entity which is subjected to the random process: is it a straight-line or two separated points?  In other terms: BQ or BQ2, or a further variant of these? Now, since Bertrand's question is usually interpreted in relation to straight lines, i.e., stick-like entities, we will first analyze BQ, then BQ2, and then also further variants.

\section{Modeling the different ``ways of tossing a stick''}

\begin{figure}[!ht]
\centering
\includegraphics[scale =0.6]{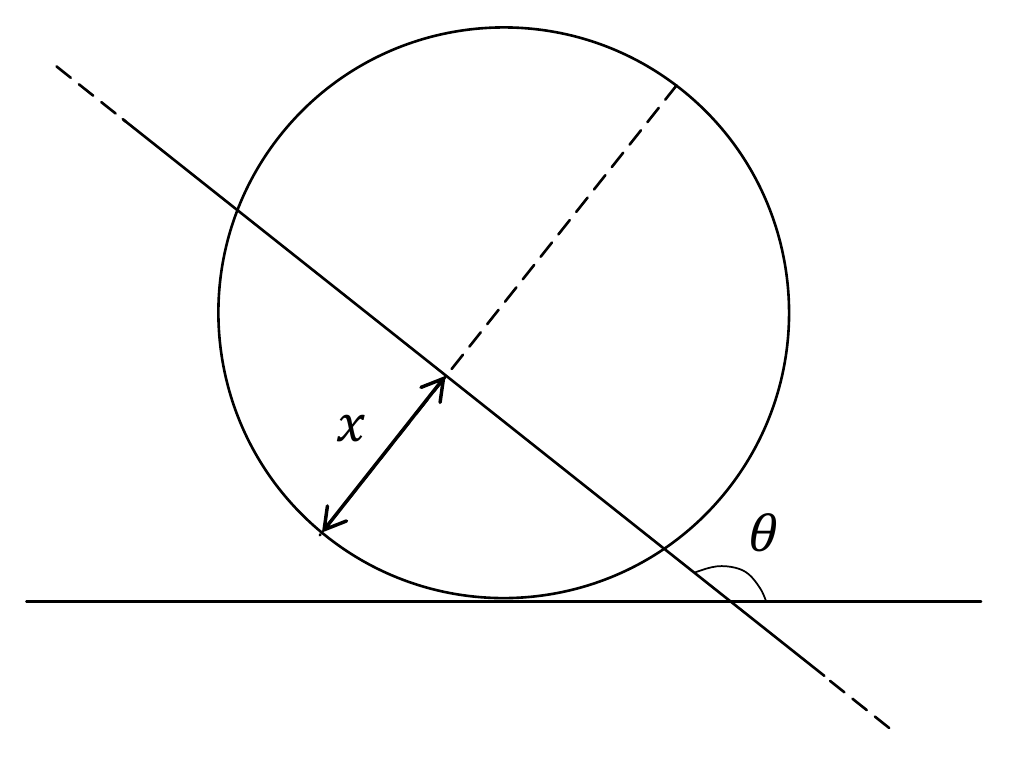}
\caption{Straight lines intersecting the unit circle can be characterized by the two parameters $\theta\in [0,\pi]$ and $x\in [0,2]$.
\label{model}}
\end{figure}

At this point,  we need to find some convenient parameters  allowing us to univocally identify a straight line intersecting the unit circle, and then give a sufficiently general description of the different ways a human being can toss a stick onto a circle. A simple parameterization, well adapted to the problem, consists in describing the different lines (sticks) in terms of the angle $\theta\in [0,\pi]$ they make with respect to a given axis, and of their distance $x\in [0,2]$ from the bottom of the circle in a direction perpendicular to their orientation, as depicted in Fig.~\ref{model}.

To modelize the different possible ``ways of tossing a stick,'' we recall that these are different in the sense that they produce different statistics of outcomes. Therefore, it is natural to characterize them by means of different probability densities $\rho(x,\theta)\geq 0$, $\int_0^2 dx\int_0^\pi d\theta \rho(x,\theta) =1$. More precisely, if $A\subset [0,2]$, and $B\subset  [0,\pi]$, then
\begin{equation}
{\cal P}(x\in A, \theta\in B|\rho) =   \int_0^2 \! \! dx  \!\int_0^\pi   \!\! d\theta\,  \chi_A(x)\chi_B(\theta) \rho(x,\theta)
\label{PAB}
\end{equation}
is the probability that, when the human operator tosses the stick in the $\rho$-way, a stick with angle in  $A$ and distance in $B$ will be obtained, where $\chi_A(x)$ and $\chi_B(\theta)$ denote the characteristic functions of intervals $A$ and $B$, respectively [$\chi_A(x)=1$, if $x\in A$, and zero otherwise, and similarly for $\chi_B(\theta)$].

Thanks to the above modelization, and considering that the length of a chord only depends on $x$, and not on $\theta$, the probability of obtaining a chord of length $L$ greater than $\sqrt{3}$, given that the stick is tossed in the  $\rho$-way, is:
\begin{eqnarray}
\lefteqn{{\cal P}(L>\sqrt{3}|\rho) =  {\cal P}\left(x\in \left[{1\over 2},{3\over 2}\right]\bigg|\rho\right)}\\
&=& \int_0^2 \!\! dx\int_0^\pi \!\! d\theta\,  \chi_{[{1\over 2},{3\over 2}]}(x)\rho(x,\theta) = \int_0^2 \!\! dx  \chi_{[{1\over 2},{3\over 2}]}(x)\rho(x),\nonumber
\label{PL3}
\end{eqnarray}
where we have defined the one-variable probability density: $\rho(x)\equiv \int_0^\pi d\theta\,\rho(x,\theta)$.

Consider now, as a simple example, the one-parameter family of probability densities:
\begin{eqnarray}
\rho_p(x) = p\, \chi_{[{1\over 2},{3\over 2}]}(x) + (1-p)\, \chi_{[0, {1\over 2}]\cup [{3\over 2},2]}(x),
\label{p}
\end{eqnarray}
where $p\in[0,1]$. Clearly, $\int_0^2 dx\rho_p(x) =1$, i.e., $\rho_p(x)$ is duly normalized, for all $p$, and we have: 
\begin{eqnarray}
{\cal P}(L>\sqrt{3}|\rho_p) = \int_0^2 dx,  \chi_{[{1\over 2},{3\over 2}]}(x)\rho_p(x) = p.
\label{PL3-bis}
\end{eqnarray}
This means that, even if we restrict the possible ways of tossing a stick to those described by the simple one-parameter family (\ref{p}), we can already generate all possible probabilities for an outcome chord to be longer that $\sqrt{3}$. And in particular, for $p={1\over 3}, {1\over 2}, {1\over 4}$, we recover the  three answers proposed by Bertrand. 

Of course, as we discussed, none of these different possible solutions answer the EQ (and therefore BQ), as the question does not indicates a specific ``way of tossing the stick,'' so that all possible ways should be considered and, according to Laplace's principle of insufficient reason, \emph{equally} considered.

\section{Defining the universal average}

The problem with which we are confronted is that of defining and calculating a uniform average over all possible $\rho(x)$, which we will call a \emph{universal average}. If we are successful in determining such huge average, then we will have provided an answer to BQ. More precisely, let us write: 
\begin{equation}
{\cal P}(L>\sqrt{3}) = \left\langle{\cal P}(L>\sqrt{3}|\rho) \right\rangle^{\!{\rm univ}}\mkern-18mu\!\!\! ,
\label{Pmean}
\end{equation}
where $\langle\,\cdot\,\rangle^{\rm univ}$ denotes the universal average, i.e., the uniform average over all possible $\rho(x)$.  What we have called the ``hard problem of Bertrand'a paradox'' is precisely that of giving a clear definition of the above universal average, and calculating its  value. 

To be able to define a uniform average over all possible uncountable $\rho(x)$, without being confronted with insurmountable technical problems  related to the foundations of mathematics, we will follow a strategy which is similar to what is done in the definition of the \emph{Wiener measure}, which is a  probability law on the space of continuous functions, describing for instance the Brownian motions. As is known, the Wiener measure is capable of attributing probabilities to continuous-time random walks, and it can do so by considering them as a limit of discrete-time processes. More precisely, in the analysis of Brownian processes, one starts with the description of particles that can move only on regular cellular structures (regular lattices), which at each step can jump from one location to another, according to a given probability law, and then one considers the (continuous) limit where these steps become infinitesimal.\cite{Martin}

To define the average (\ref{Pmean}), we will proceed with a similar logic, first considering discretized (cellular) probability densities $\rho_{n}(x)$, whose uniform average is perfectly well-defined,  then  taking the infinite limit $n\to\infty$. To show that such procedure is consistent, in the sense that it allows to include in the average all possible probability densities, we have to show that all $\rho(x)$ (including generalized functions) can be approximated by discretized (cellular) probability densities $\rho_n(x)$, in the sense that, given a $\rho(x)$, and an interval $A\subset [0,2]$, we can always find a suitable sequence of cellular  $\rho_{n}(x)$ such that ${\cal P}(x\in A|\rho_n) \to {\cal P}(x\in A|\rho)$, as $n\to\infty$, for all $A$. 

To see this,  we partition the interval $[0, 2]$ into $n=m\ell$ elementary intervals: 
\begin{equation}
\left[0,{2\over n}\right], \left[{2\over n},{4\over n}\right],\dots,\left[{2(j-1)\over n},{2j\over n}\right],\dots,\left[2-{2\over n},2\right].\nonumber
\end{equation} 
These elementary intervals (one-dimensional elementary cells), of length ${2\over n}$, are in turn contained in $m={n\over \ell}$ larger intervals, of length ${2\over m} = {2\ell\over n}$, which are the following:
\begin{equation}
\left[0,{2\over m}\right], \left[{2\over m},{4\over m}\right],\dots, \left[ {2(i-1)\over m}, {2i\over m}\right],\dots, \left[2- {2\over m},2\right].\nonumber
\end{equation} 
In other terms, each cell $S_i \equiv [ {2(i-1)\over m}, {2i\over m}]$, is made of $\ell$ elementary cells $\sigma_{i,j} \equiv [ {2(i-1)\over m}+{2(j-1)\over n},{2(i -1)\over m}+{2j\over n}]$, i.e., $S_i=\cup_{j=1}^{\ell} \sigma_{i,j}$, and $[0,2] =\cup_{i=1}^{m} S_i = \cup_{i=1}^{m}\cup_{j=1}^{\ell} \sigma_{i,j}$.

Let $A\equiv[x_1,x_2]\subset [0,2]$, and assume that $x_1$ and $x_2$ are such that $x_1\in ({2(j-1)\over m}, {2j\over m}]$ and $x_2\in ({2(k-1)\over m}, {2k\over m}]$, for some given integers $j$ and $k$. These two conditions can also be expressed as ${mx_1\over 2}\in (j-1,j]$ and ${mx_2\over 2}\in (k-1,k]$, $k\geq j$. By definition of the \emph{ceiling function}, we thus have $\lceil {mx_1\over 2}\rceil=j$ and  $\lceil {mx_2\over 2}\rceil=k$, and we can write: 
\begin{eqnarray}
\label{partition4}
\lefteqn{{\cal P}(x\in [x_1,x_2]|\rho)= \int_{0}^{2} \chi_{[x_1,x_2]}(x) \rho(x) dx}\\
&&=\int_{x_1}^{x_2} \!\!\!\rho(x) dx =\!\!\! \sum_{i=\lceil {mx_1\over 2}\rceil +1}^{\lceil {mx_2\over 2}\rceil -1}\int_{S_i} \rho(x)dx +r_{m}(x_1,x_2|\rho),\nonumber
\end{eqnarray}
where the rest
\begin{equation}
\label{restrho}
r_{m}(x_1,x_2|\rho) = \int_{x_1}^{\lceil {mx_1\over 2}\rceil {2\over m}}\rho(x)dx + \int_{(\lceil {mx_2\over 2}\rceil-1) {2\over m}}^{x_2}\rho(x)dx
\end{equation}
tends to zero, as $m\to\infty$, considering that 
\begin{equation}
\label{ceilinglimit}
\lim_{n\to\infty}{\lceil nx\rceil \over n}=x.
\end{equation}

\begin{figure*}[!ht]
\centering
\includegraphics[scale =.9]{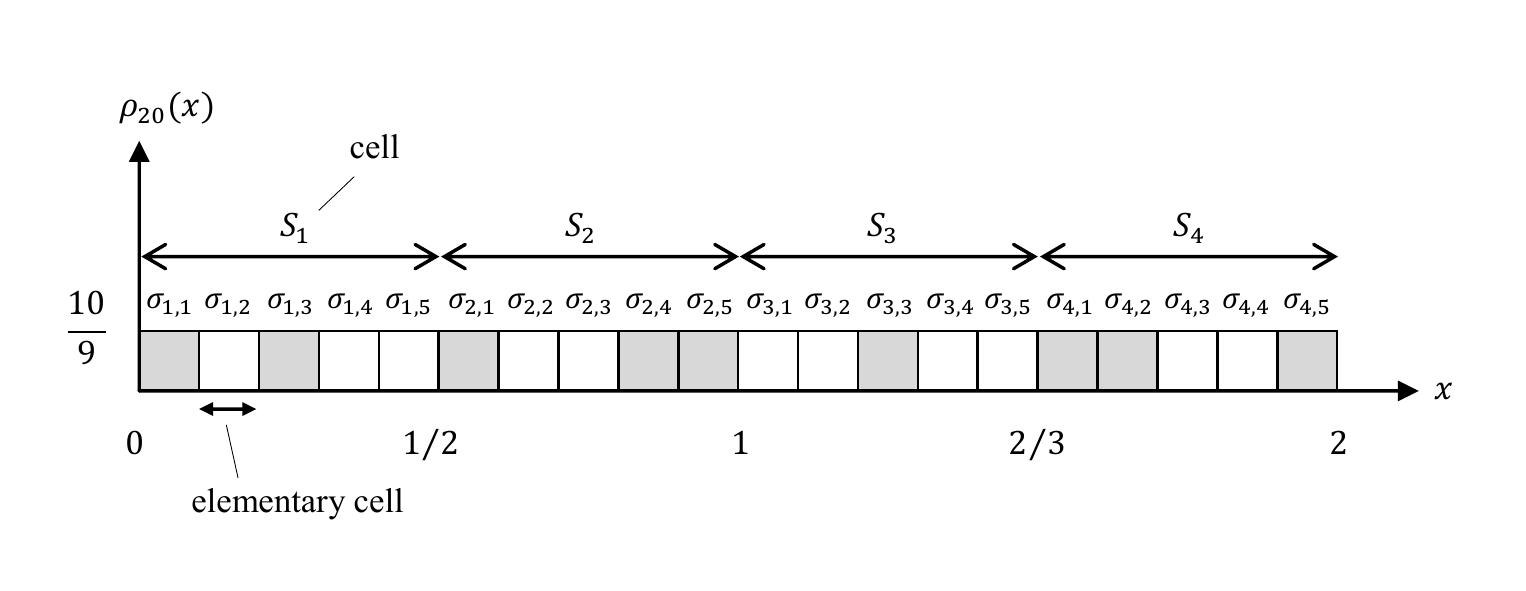}
\caption{A graphical representation of a cellular probability density made of $n=20$ elementary cells. The $n-\tilde{n} =11$ white squares denote the elementary cells in which the probability density is zero, whereas the $\tilde{n} =9$ grey squares denote those in which it takes a constant value, equal to ${10\over 9}$. Since each elementary cell has width ${1\over 10}$, we clearly have $\int_0^2 dx \rho_{20}(x) =  9 {10\over 9}{1\over 10}=1$.
\label{cells}}
\end{figure*}

At this point, we define the cellular probability density:
\begin{equation}
\label{cellular}
\rho_{n}(x)={n\over 2\tilde{n}}\chi_{m\ell}(x),
\end{equation}
where $n\equiv m\ell$, $\chi_{m\ell}(x)$ denotes a  step-like function taking only the constant values $1$ or $0$ inside each elementary cell $\sigma_{i,j}$, and $\tilde{n}$ is the total number of elementary cells in which $\chi_{m\ell}(x)$ takes the value $1$ (see Fig.~\ref{cells}).
For a cellular probability density of this kind, (\ref{partition4})  becomes:
\begin{eqnarray}
\label{partition5}
\lefteqn{{\cal P}(x\in [x_1,x_2]|\rho_n) =\int_{x_1}^{x_2} \rho_n(x) dx}\nonumber\\
&=&\sum_{i=\lceil {mx_1\over 2}\rceil +1}^{\lceil {mx_2\over 2}\rceil -1}{\tilde{n}_{i}\over \tilde{n}}+r_{m}(x_1,x_2|\rho_n),
\end{eqnarray}
where $\tilde{n}_{i}$ is the number of elementary cells of $S_i$ in which $\chi_{m\ell}(x)$ takes the value $1$. Comparing (\ref{partition4}) with (\ref{partition5}), we obtain:
\begin{equation}
{\cal P}(x\in [x_1,x_2]|\rho) -{\cal P}(x\in [x_1,x_2]|\rho_n) = \sum_{i=\lceil {mx_1\over 2}\rceil +1}^{\lceil {mx_2\over 2}\rceil -1}\left(\int_{S_i} \rho(x)dx -{\tilde{n}_{i}\over \tilde{n} }\right)+[r_{m}(x_1,x_2|\rho)-r_{m}(x_1,x_2|\rho_n)].
\label{difference-m-l}
\end{equation}

Clearly, we can always choose $\rho_{n}(x)$ in such a way that ${\tilde{n}_{i}\over \tilde{n}}\to\int_{S_i} \rho(x)dx$, as $\ell\to\infty$, for all $i=1,\dots,m$. This because, for each $i$, the probability $\int_{S_i} \rho(x)dx$ is a real number with value in the interval $[0,1]$, and rational numbers of the form ${\tilde{n}_{i}\over \tilde{n}}$, with $0\leq \tilde{n}_{i}\leq \tilde{n}$, $\tilde{n}>0$, are dense in $[0,1]$. Therefore, for such a choice of $\rho_{n}(x)\equiv \rho_{m\ell}(x)$, by taking  the limit $\ell\to\infty$, the sum in (\ref{difference-m-l}) vanishes, and by taking the limit $m\to\infty$, also the two rests in (\ref{difference-m-l}) vanish, and we can conclude that ${\cal P}(x\in [x_1,x_2]|\rho_n) \to {\cal P}(x\in [x_1,x_2]|\rho)$, as $n\to\infty$.

In other terms, any $\rho(x)$ can be described as the limit of a suitably chosen sequence of cellular (discretized) $\rho_{n}(x)$, as the number of cells $n$ tends to infinity. Also, since a cellular probability density $\rho_{n}(x)$ is only made of a finite number $n$ of elementary cells, which can either be of the zero $(z)$ or non-zero $(\bar{z})$ kind, if we exclude the case of a $\rho_{n}(x)$ formed only by zero cells (the trivial case $\rho_{n}(x)\equiv 0$, producing no outcomes), we have that, given a  $n \in \mathbb{N}$, the total number of possible $\rho_{n}$ is $C_{n}^0 + C_{n}^1 + C_{n}^2 + \cdots + C_{n}^{n} -1 = 2^{n}-1$. Therefore, for each $n$, we can unambiguously define the average probability: 
\begin{equation}
\label{average1}
{\cal P}(x\in [x_1,x_2]|n) \equiv {1\over 2^{n}-1}  \sum_{\rho_{n}}{\cal P}(x\in [x_1,x_2]|\rho_n),
\end{equation}
where the sum runs over all the possible $2^{n}-1$ (non-zero) cellular probability densities $\rho_{n}(x)$, made of $n$ elementary cells. 

Clearly, ${\cal P}(x\in [x_1,x_2]|n)$ is the probability that the distance $x$ for the stick will be found in the interval $[x_1,x_2]$, when the human operator randomly chooses one among the $2^{n}-1$ possible ``$\rho_n$-ways of tossing the stick,'' and then tosses it accordingly. It is worth observing that the uniform average (\ref{average1}) being over a finite number of possibilities, it is uniquely defined and doesn't suffer from possible ambiguities. Also, considering that the $\rho_{n}(x)$ are dense in the space of  probability densities (in the sense specified above), we are now in a position to give the following general definition of a universal average:
\begin{equation}
\left\langle{\cal P}(x\in [x_1,x_2]|\rho) \right\rangle^{\rm univ} \equiv  \lim_{n\to\infty}{\cal P}(x\in [x_1,x_2]|n),
\label{universaldefinition}
\end{equation}
where ${\cal P}(x\in [x_1,x_2]|n)$ is the arithmetic mean (\ref{average1}).

\section{Solving the ``hard'' problem}

We can now state and prove the following theorem:
\\

\noindent {\bf Theorem}. \emph{The universal average (\ref{universaldefinition}), over all possible probability densities $\rho(x)$, on the interval $[0,2]$, produces the same  probabilities as those calculated by means of a uniform probability density $\rho_u(x) ={1\over 2}$, in the sense of the equality:
\begin{equation}
\label{theoremuniversal}
\left\langle{\cal P}(x\in [x_1,x_2]|\rho) \right\rangle^{\rm univ} ={\cal P}(x\in [x_1,x_2]|\rho_u)= {x_2-x_1\over 2}.
\end{equation}
}

\noindent Considering that Bertrand's condition corresponds to the choice $x_1={1\over 2}$ and $x_2={3\over 2}$, we find that the  answer to BQ is: 
\begin{equation}
{\cal P}(L>\sqrt{3}) =  \left\langle  {\cal P}\left(x\in \left[{1\over 2},{3\over 2}\right]|\rho\right) \right\rangle^{\rm univ} \!\!\!\!\!\! = {1\over 2}.
\label{solution}
\end{equation}

To prove the theorem, it is sufficient to show that for all $n\in \mathbb{N}$:
\begin{equation}
\label{averagerho2}
{\cal P}(x\in [x_1,x_2]|n) ={\cal P}(x\in [x_1,x_2]|\rho_{u;n}),
\end{equation}
where $\rho_{u;n}(x)$ designates a cellular uniform probability density made of $n$ non-zero elementary cells. Then, observing that ${\cal P}(x\in [x_1,x_2]|\rho_{u;n})\to {\cal P}(x\in [x_1,x_2]|\rho_{u})$, as $n\to\infty$, (\ref{averagerho2})  clearly proves (\ref{theoremuniversal}). For simplicity, we limit our discussion to intervals of the form: $A_n(i,j)\equiv \left[i{2\over n},j{2\over n}\right]$, with $i= \lceil n {x_1\over 2}\rceil$ and $j= \lceil n {x_2\over 2}\rceil$, $0\leq i\leq j\leq n$. This will be sufficient, as is clear that, in view of (\ref{ceilinglimit}), $A_n(i,j)\to [x_1,x_2]$, as $n\to\infty$. 

\begin{figure}[!ht]
\centering
\includegraphics[scale =0.6]{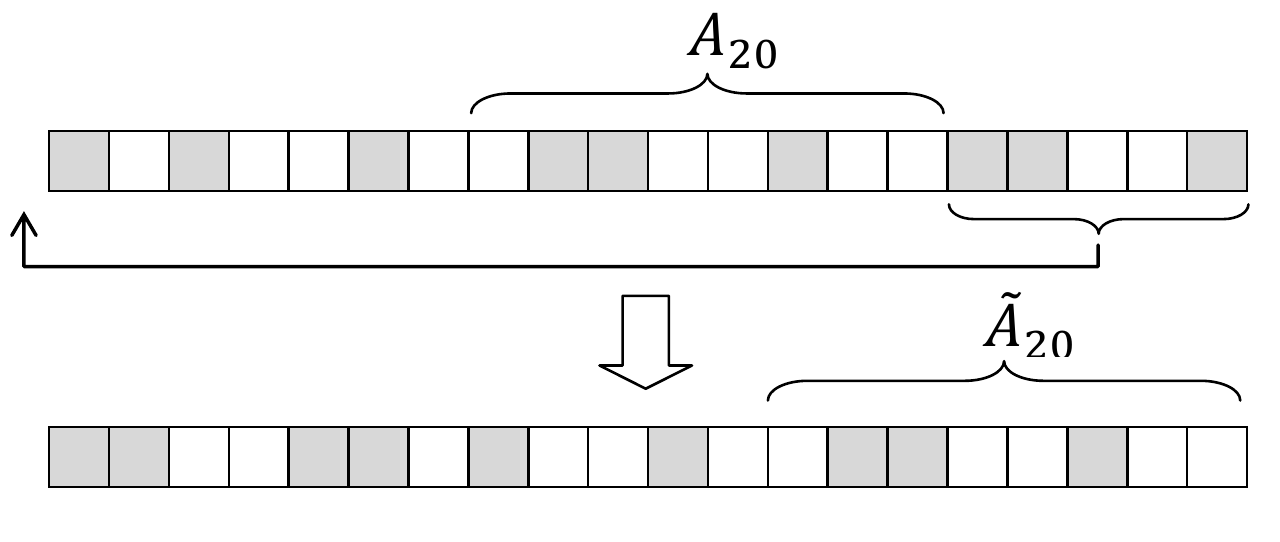}
\caption{An example of cells' displacement producing $\tilde\rho_{n}(x)$ from $\rho_{n}(x)$, here in the case of an interval $A_n(i,j)$ corresponding to $i=7$, $j=15$ and $n=20$.
\label{cellsdisplacement}}
\end{figure}

What we need to calculate is the average probability (\ref{average1}), i.e., we need to show that:
\begin{equation}
\label{averagerho2-bis}
\sum_{\rho_{n}}{\cal P}(x\in A_n(i,j)|\rho_n)=\left({2^{n}-1} \right) {\cal P}(x\in A_n(i,j)|\rho_{u;n}),
\end{equation}
for all integers $i,j,n$ such that $0\leq i\leq j\leq n$. To do so, we start by observing that to each $\rho_n(x)$ we can always associate a $\tilde\rho_n(x)$, obtained in the following way: one takes the $n-j$ elementary cells on the right side of $A_n(i,j)$ and displace them on the left side, so that the $(j+1)$-th elementary cell will now become the first one; see Fig.~\ref{cellsdisplacement}. 
This new configuration of cells defines the cellular probability density $\tilde\rho_n(x)$, and it is clear that by defining the displaced interval $\tilde{A}_n(i,j)\equiv \left[(n-(j-i)){2\over n},2\right]$, we have the equality:  ${\cal P}(x\in A_n(i,j)|\rho_n) ={\cal P}(x\in \tilde{A}_n(i,j)|\tilde\rho_n)$. In other terms, without loss of generality, we can always reason as if the interval expressing the condition on the position $x$ of the stick is placed on the right hand side of the cellular structure, i.e., we can always set $j=n$ and consider only intervals of the form: $A_n^i\equiv A_n(i,n) = \left[i{2\over n},n{2\over n}\right]$, containing $n-i$ elementary cells. 

For the uniform $\rho_{u;n}(x)$, we have ${\cal P}(x\in A_n^i|\rho_{u;n})={n-i\over n}$, as is clear that in this case the number of  non-zero elementary cells in $A_n^i$ (resp., in the entire interval $[0,2]$) is equal to the total number $n-i$ (resp., $n$) of elementary cells it contains, so that (\ref{averagerho2-bis}) becomes:
\begin{equation}
\label{averagerho2-tris}
\sum_{\rho_{n}}{\cal P}(x\in  A_n^i|\rho_n)=\left({2^{n}-1} \right) {n-i\over n}.
\end{equation}

The cases $i=0$ and $i=n$ are trivial, since ${\cal P}(x\in  A_n^0|\rho_n)=1$ and ${\cal P}(x\in  A_n^n|\rho_n)=0$, for all $n$. Thus, we only need to consider the values $i=1,\dots,n-1$. For simplicity, we adopt the following notation: ${\cal P}(i|c_1\dots c_n)\equiv {\cal P}(x\in  A_n^i|\rho_n)$, where  $(c_1\dots c_n)$, $c_j\in\{z,\bar{z}\}$, denotes the sequence of zero $(z)$  and non-zero $(\bar{z})$ elementary cells characterizing $\rho_n(x)$; see Fig.~\ref{cells}. For $i=1$, (\ref{averagerho2-tris}) becomes: 
\begin{equation}
\label{i=1}
\sum_{(c\cdots)} P(1|c\cdots) = (2^n-1){n-1\over n},
\end{equation}
and we can write:
\begin{equation}
\sum_{(c\cdots)} P(1|c\cdots) =\sum_{(z\cdots)} P(1|z\cdots)+ \sum_{(\bar{z}\cdots )} P(1|\bar{z}\cdots ),
\end{equation}
where the first sum in the r.h.s. of the equation runs over all $\rho_n(x)$ starting with a left zero $(z)$ elementary cell, and the second sum runs over all $\rho_n(x)$ starting with a left non-zero $(\bar{z})$ elementary cell. We can observe that all probabilities in the first sum are equal to $1$, so that the sum is equal to $2^{n-1}-1$. Also, the second sum can be written as $\sum_{k=0}^{n-1} {k\over k+1} {n-1 \choose k}$. Using a symbolic computational program (like Mathematica, of Wolfram Research, Inc.), one obtains the exact identity:
\begin{equation}
\label{Wolfram}
\sum_{k=0}^{n} {k\over k+1} {n \choose k} = {2^n (n-1) +1\over n+1},
\end{equation}
so that:
\begin{eqnarray}
\sum_{(c\cdots)} P(1|c\cdots) &=& 2^{n-1}-1 +  {2^{n-1} (n-2) +1\over n}\nonumber\\
&=& (2^n-1){n-1\over n},
\end{eqnarray}
which proves (\ref{i=1}). 

To prove (\ref{averagerho2-tris}) for all $i\in\{1,\dots,n-1\}$, we reason by recurrence. We have shown that the equality holds for $i=1$; let us assume it holds for some $i$, and that this implies it also holds for $i+1$. We write:
\begin{eqnarray}
\label{sumgeneral}
\lefteqn{\sum_{(c \cdots)} P(i+1|c\cdots) =}\\ 
&&\sum_{(\cdots z\cdots)} P(i+1|\cdots z\cdots)
+ \sum_{(\cdots \bar{z}\cdots )} P(i+1|\cdots \bar{z}\cdots ),\nonumber
\end{eqnarray}
where the first sum, in the r.h.s. of the equation, runs over all $\rho_n(x)$ having a zero $(i+1)$-th elementary cell, and the second sum  runs over all $\rho_n(x)$ having a non-zero $(i+1)$-th elementary cell. Observing that $P(i+1|\cdots z\cdots)=P(i|\cdots z\cdots)$, we can write for the first sum:
\begin{eqnarray}
\label{sum-final}
\sum_{(\cdots z\cdots)} P(i+1|\cdots z\cdots)&=&\sum_{(\cdots z\cdots)} P(i|\cdots z\cdots) = \sum_{(\cdots z\cdots)} P(i|\cdots z\cdots)+\sum_{(\cdots \bar{z}\cdots)} P(i|\cdots \bar{z}\cdots)-\sum_{(\cdots \bar{z}\cdots)} P(i|\cdots \bar{z}\cdots)\nonumber\\
&=&\sum_{(c\cdots)} P(i|c\cdots)-\sum_{(\cdots \bar{z}\cdots)} P(i|\cdots \bar{z}\cdots) =  (2^n-1){n-i\over n}-\sum_{(\cdots \bar{z}\cdots)} P(i|\cdots \bar{z}\cdots),
\end{eqnarray}
where for the third equality we have added and subtracted the same quantity, and for the last equality we have used (\ref{averagerho2-tris}) and the recurrence hypothesis. Then, (\ref{sumgeneral}) becomes:
\begin{eqnarray}
\label{sum-general2}
\lefteqn{\sum_{(c \cdots)} P(i+1|c\cdots) =(2^n-1){n-i\over n}}\nonumber\\
&&+\sum_{(\cdots \bar{z}\cdots)} \left[ P(i+1|\cdots \bar{z}\cdots) - P(i|\cdots \bar{z}\cdots)\right].
\label{difference}
\end{eqnarray}

Denoting $k_i$ the number of non-zero elementary cells at the right of the $i$-th cell, and $k$ the total number of non-zero elementary cells, for a cellular probability density of the $(\cdots \bar{z}\cdots)$ kind, we have $P(i|\cdots \bar{z}\cdots)={k_i\over k}$, and $P(i+1|\cdots \bar{z}\cdots)={k_i-1\over k}$, so that the difference of probabilities in (\ref{difference}) is equal to $-{1\over k}$, and is independent of $k_i$. Using the exact identity (which again can be obtained using a symbolic computational program, like Mathematica, of Wolfram Research, Inc.):
\begin{equation}
\label{Wolfram3}
\sum_{k=0}^{n} {1\over k+1} {n \choose k}={2^{n+1}-1\over n+1},
\end{equation}
we therefore obtain
\begin{equation}
\label{sumgeneral3}
-\sum_{(\cdots \bar{z}\cdots)} {1\over k(\cdots \bar{z}\cdots)} =-\sum_{k=0}^{n-1} {1\over k+1} {n-1 \choose k}=-{2^{n}-1\over n},
\end{equation}
and inserting (\ref{sumgeneral3}) into (\ref{sum-general2}), we finally obtain:
\begin{eqnarray}
\sum_{(c \cdots)} P(i+1|c\cdots) &=&(2^n-1){n-i\over n}-{2^{n}-1\over n}\\
&=&(2^n-1){n-(i+1)\over n},\label{final}
\end{eqnarray}
which proves that (\ref{averagerho2-tris}) also holds for $i+1$, thus completing the recurrence proof.

\section{The modified Bertrand's questions}

In the previous section we have proved a theorem showing that the uniform average over all possible ways of tossing a stick produces the answer ${\cal P}(L>\sqrt{3}) =  {1\over 2}$, to the original BQ. What about the modified version BQ2? Can we apply the notion of universal average to also solve BQ2? 

\begin{figure}[!ht]
\centering
\includegraphics[scale =0.65]{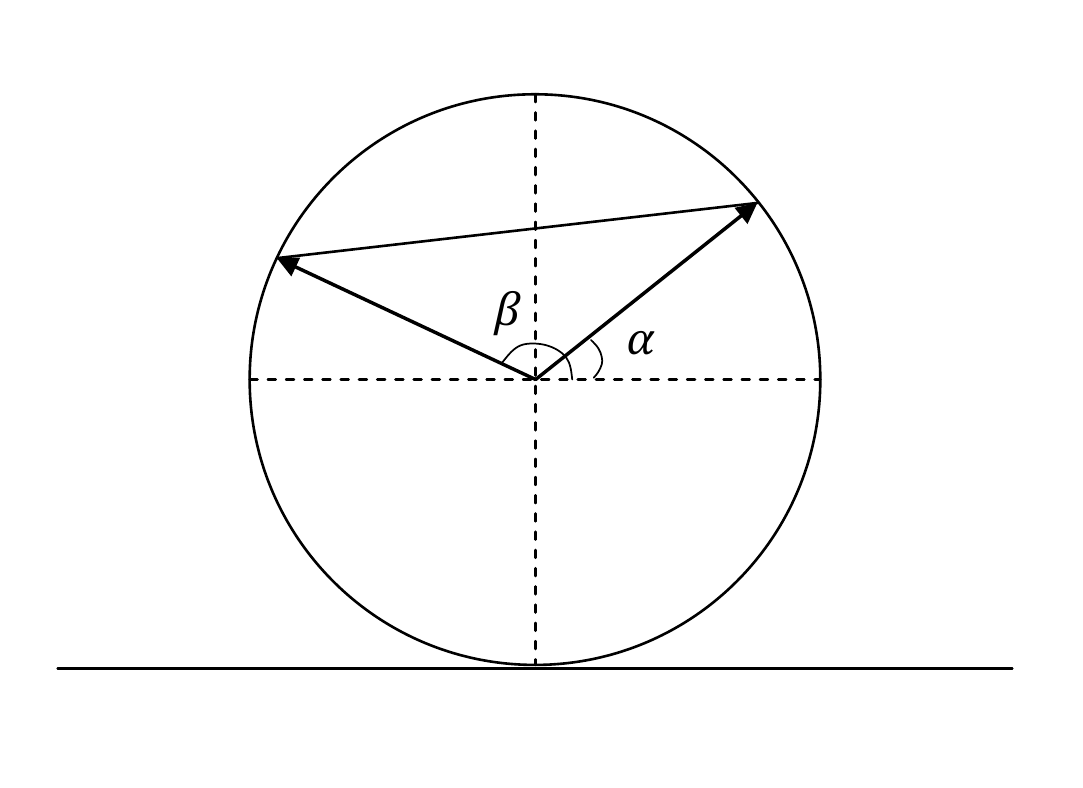}
\caption{Two points on the circle (i.e., on the border of the disk) are parameterized by their two angles $\alpha,\beta\in [0,2\pi]$, whose difference determines the length of the associated chord.
\label{model-twopoints}}
\end{figure}

\begin{figure}[!ht]
\centering
\includegraphics[scale =0.55]{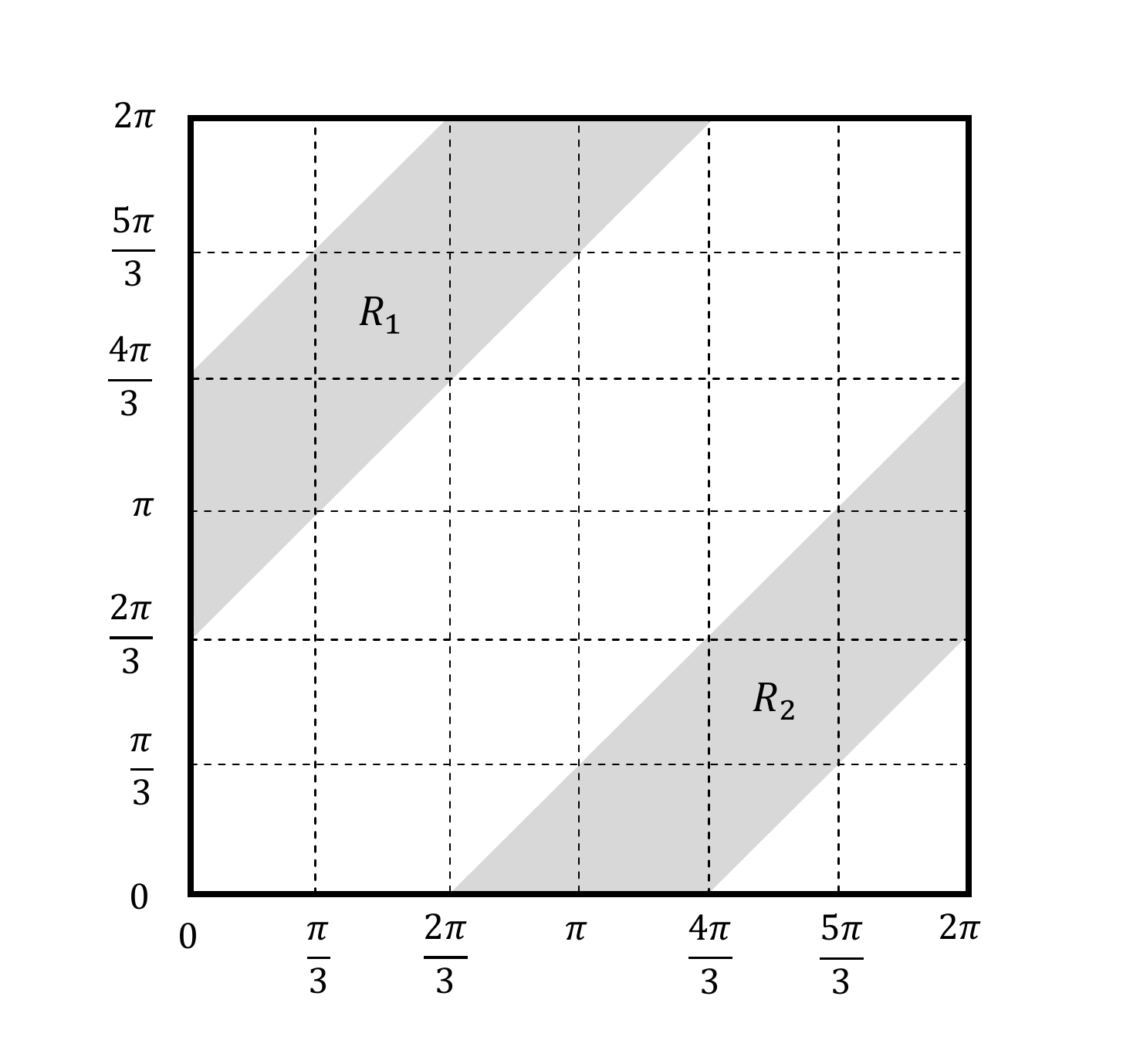}
\caption{The Cartesian square $[0,2\pi]\times [0,2\pi]$, of total area $4\pi^2$, representing all possible couples of points $(\alpha,\beta)$ on a unit circle. The grey region $R=R_1\cup R_2$, of area ${4\pi^2 \over 3}$, represents the couples of points on the circle giving rise to a chord of length  $L>\sqrt{3}$.
\label{square}}
\end{figure}

For this, as we did for the sticks, we need  to choose some convenient parameters to univocally identify two points intersecting the unit circle (representing the two peebles in EQ2). A natural and simple way to do this is by means of two angles $(\alpha,\beta)\in [0,2\pi]\times [0,2\pi]$; see Fig.~\ref{model-twopoints}.
Then,  the probability density modeling the different possible ``ways of tossing two pebbles'' onto the circle's curve will be a function of these two angles, i.e., $\rho\equiv  \rho (\alpha,\beta)$, and  Bertrand's condition $L>\sqrt{3}$ is satisfied when ${2\pi\over 3}\leq |\beta -\alpha|\leq {4\pi\over 3}$. This condition corresponds to the grey region $R=R_1\cup R_2$, represented in Fig.~\ref{square}, which has a total area of ${4\pi^2 \over 3}$ (the area is easily calculated if one observes that $R$ is made by $12$ small squares of side ${\pi\over 3}$; see Fig.~\ref{square}).  
Therefore, denoting $\chi_R(\alpha,\beta)$ the characteristic function of region $R$, the probability that by tossing two pebbles in the $\rho$-way, they will generate a chord of length $L>\sqrt{3}$ (when they both intersect the circle), is given by:
\begin{equation}
{\cal P}((\alpha,\beta)\in R|\rho) =   \int_0^{2\pi} \!\! d\alpha \int_0^{2\pi} \!\! d\beta\,  \chi_R(\alpha,\beta) \rho (\alpha,\beta).
\label{Palfabeta}
\end{equation}

As a simple example, consider the following one-parameter family of probability densities:
\begin{equation}
\rho_p(\alpha,\beta) = p\, \chi_{R}(\alpha,\beta) + (1-p)\, \chi_{\bar{R}}(\alpha,\beta),
\label{pbis}
\end{equation}
where $p\in[0,1]$, and $\bar{R} = [0,2\pi]\times [0,2\pi] \setminus R$. Clearly, ${\cal P}((\alpha,\beta)\in R|\rho_p) =p$, for all $p$, so that also in this case, even when restricting the ways of tossing the two pebbles to those described by (\ref{pbis}), we can already generate all possible values for the probability that the final chord has length $L>\sqrt{3}$. 

But again, none of these different possible solutions answer EQ2, and therefore BQ2, as the question does not specify a specific ``way of tossing the two peebles,'' so that all possible ways should be considered, and according to Laplace's principle of insufficient reason they should be given equal a priori weight. In other terms, we need to consider a universal average over all possible $\rho (\alpha,\beta)$-ways of tossing the two pebbles.

Differently from the previous analysis, instead of the one-dimensional interval $[0,2]$, we now have the two-dimensional square $[0,2\pi]\times [0,2\pi]$. However, the proof can be straightforwardly generalized. Indeed, instead of tessellating a one-dimensional interval by means of one-dimensional cells, we have now to tessellate a two-dimensional square by means of two-dimensional square cells. It is then always possible to devise a method to enumerate these square cells and order them in a convenient way along a line, thus transforming the two-dimensional problem into a one-dimensional one (see Ref.~\onlinecite{AertsSassoli} for the details). In this way, a similar proof as per above applies, and one can show that, for any region $R\subset [0,2\pi]\times [0,2\pi]$:
\begin{equation}
\label{theoremuniversal2}
\left\langle{\cal P}((\alpha,\beta)\in R|\rho) \right\rangle^{\rm univ} ={\cal P}((\alpha,\beta)\in R|\rho_u)= {\mu(R)\over 4\pi^2},
\end{equation}
where $\rho_u(\alpha,\beta)={1\over 4\pi^2}$, and $\mu(R)$ denotes the Lebesgue measure of region $R$. If $R$ corresponds to the above mentioned region, characterizing Bertrand's condition, then $\mu(R)={4\pi^2 \over 3}$, and one finds that the unique answer to BQ2 is:
\begin{equation}
{\cal P}(L>\sqrt{3}) =  \left\langle  {\cal P}\left((\alpha,\beta)\in R |\rho\right) \right\rangle^{\rm univ} = {1\over 3}.
\label{solution2}
\end{equation}

For completeness, we briefly describe a third variant of BQ, which is the following:\\

\noindent {\bf  Modified Bertrand's Question (BQ3)}: We draw \emph{at random} a  point inside a circle and a chord having such point as its middle point. What is the probability that it is longer than the side of the inscribed equilateral triangle?
\\

\noindent Clearly, this  version of Bertrand's question does not admit a physical realization in terms of a single stick, nor of a pair of pebbles. Indeed, both a point and a straight line are now the entities which are subjected to the randomization. A convenient physical realization of this additional modified question is therefore, for instance, the following:\\

\noindent {\bf Experimental Realization of BQ3 (EQ3)}: What is the probability that a \emph{human being}, by \emph{tossing} a pebble and a stick inside a unit circle drawn on the floor, when bringing the middle point of the stick at the position of the pebble (without altering its orientation), will define on it a chord of length $L$ greater than $\sqrt{3}$?\\

Obviously, the orientation of the stick doesn't influence the length of the obtained chord, so that the probability is actually independent of the tossing of the stick. This means that only the different ``ways of tossing a pebble'' inside the circle will have to be considered. The position of the pebble inside the circle is determined by a point $(x,y)$ in the unit disk $D_1=\{(x,y)\in \mathbb{R}^2: x^2+y^2\leq 1\}$, and Bertrand's condition is met by a pebble such that $(x,y)\in D_{1\over 2}$, where $D_{1\over 2}=\{(x,y)\in \mathbb{R}^2: x^2+y^2\leq {1\over 2}\}$ is the internal disk of radius ${1\over 2}$, depicted in Fig.~\ref{circle} (c). 

Thus, denoting by $\chi_{D_{1\over 2}}(x,y)$ the characteristic function of the disk $D_{1\over 2}$, the probability that by tossing a single pebble in the $\rho$-way, this will generated a chord (having the pebble as its middle point) of length $L>\sqrt{3}$, is given by:
\begin{equation}
{\cal P}((x,y)\in D_{1\over 2}|\rho) =    \iint_{D_1} dx \,dy \,  \chi_{D_{1\over 2}}(x,y) \rho (x,y),
\label{Pxy}
\end{equation}
and considering once more the simple example of a probability density of the form:
\begin{equation}
\rho_p(x,y) = p\, \chi_{D_{1\over 2}}(x,y) + (1-p)\, \chi_{\bar{D}_{1\over 2}}(x,y),
\label{ptris}
\end{equation}
where $p\in[0,1]$, and $\bar{D}_{1\over 2} = D_1 \setminus D_{1\over 2}$, we have ${\cal P}((x,y)\in D_{1\over 2}|\rho_p) =p$, so that all possible numerical values for the probability can be easily obtained. 

As we did for the previous questions, to answer EQ3, and therefore BQ3, we need to consider the universal average over all possible $\rho (x,y)$-ways of tossing the pebble. One obtains in this way that, for any region $D\subset D_1$ (see Refs.~\onlinecite{AertsSassoliOne, AertsSassoli} for the method):
\begin{equation}
\label{theoremuniversal2}
\left\langle{\cal P}((x,y)\in D|\rho) \right\rangle^{\rm univ} ={\cal P}((x,y)\in D|\rho_u)= {\mu(D)\over \pi},
\end{equation}
where $\rho_u(\alpha,\beta)={1\over \pi}$, and $\mu(D)$ denotes the Lebesgue measure of region $D$. If $D$ corresponds to $D_{1\over 2}$, then $\mu(D_{1\over 2})={\pi\over 4}$, and we find that the unique answer to BQ3 is:
\begin{equation}
{\cal P}(L>\sqrt{3}) =  \left\langle  {\cal P}\left((x,y)\in D_{1\over 2} |\rho\right) \right\rangle^{\rm univ} \!\!\!\!= {1\over 4}.
\label{solution3}
\end{equation}

\section{Discussion}

In his interesting analysis of Bertrand's paradox, Shackel  pointed out that two different strategies have been adopted so far, for resolving the paradox, which he calls the ``distinction strategy'' and the ``well-posing strategy,''~\cite{Shackel} indicating Marinoff~\cite{Marinoff} and Jaynes~\cite{Jaynes} as the two authoritative representatives of these two strategies, respectively.  The approach we have here presented, however, cannot be reduced to one of them. 

Indeed, our analysis shows that both strategies need to be adopted, to properly answer BQ, and its possible variants, as the question actually contains two distinct problems, which we have metaphorically called the ``easy'' and ``hard'' problem. Only when the ``easy'' problem is  addressed, by operating a \emph{distinction} between non equivalent questions,  the ``hard'' problem becomes \emph{well-posed}, and therefore solvable. 

Marinoff, in his resolution of Bertrand's paradox,~\cite{Marinoff} rightly emphasizes, as we did, that BQ is vaguely posed. However, the important difference between Marinoff's approach, and our approach, is that he identifies the primary source of ambiguity in a lack of sufficient specification of the random process selecting the chords, whereas we identify the primary source of ambiguity in a lack of sufficient specification of the nature of the entity which is subjected to the random process. To put it another way, it is not the term ``at random'' which is the source of ambiguity in the BQ, but the nature of the entity which is randomized. Is such entity a straight line (a stick), a couple of points (two pebbles), or a point and a straight line (a pebble and a stick)? 

Only once the above question is clearly answered, by specifying the nature of the entity which is subjected to the random process, one obtains a well-posed problem, which can then be solved. The well-posed problem we obtain, however, is different from the well-posed problem considered by Jaynes, although we find the same numerical answer. Jaynes considers that what is left unspecified in the problem is the size, orientation and location of the circle, and that this requires the solution to be scale, rotation and translation invariant, and that only one solution can meet this requirement (actually, the requirement of translation invariance is already sufficiently strong to determine the result uniquely~\cite{Jaynes}).  

Differently from Jaynes, we considered that what is left unspecified in the problem, at a more fundamental level, is the ``way the different chords are selected,'' that is, the ``way the randomization is obtained.'' This means that what we primarily lack knowledge about, is not which chord is each time selected, but the ``way it is selected.'' In other terms, what we have to consider is not a uniform average over all the possible chords, but a uniform average over all possible ``ways of selecting a chord.'' And when such huge average, which we have called a \emph{universal average}, is defined in a mathematically consistent way, one obtains the unique answer $(\ref{solution})$.

Therefore, BQ can be consistently answered by applying what Shackel calls the \emph{principle of metaindifference}. Quoting from Ref.~\onlinecite{Shackel}: ``Ignorance of which method of random choice has been used is just more ignorance, and the principle of indifference is supposed to warrant applying equiprobability over equal ignorance, so equiprobability should be assigned to \emph{those} possibilities. I called this \emph{metaindifference}.'' 

However, in his analysis Shackel considers that metaindifference fails to solve Bertrand's paradox, although he also prudently concedes that:~\cite{Shackel} ``Perhaps there are many ways to be metaindifferent, and perhaps a variety of notions of metaindifference could usefully be developed. Among those there might be some which avoid the arguments I shall shortly make. The question would then be whether they do any useful work in addressing Bertrand's paradox.''

We think that our analysis precisely avoids the arguments presented by Shackel, as it doesn't lead to the vicious regress he indicates (see Ref.~\onlinecite{Shackel}, Sec. 12). This because, on one hand, we have disambiguated Bertrand's question by specifying the nature of the randomized entity, and on the other hand we have identified where and how metaindifference should be applied in the problem, giving rise to a \emph{universal average} which can be calculated by a discretization procedure, much in the spirit of what is done in mathematics with the so-called Wiener process. 

Therefore, our conclusion is that Bertrand's paradox should not any more be considered a paradox, and to stand in refutation of the Laplacean principle of insufficient reason.  

It could be objected that in our approach Bertrand's paradox recurs in a different form, as we have obtained three different averages -- (\ref{solution}), (\ref{solution2}) and (\ref{solution3}) -- which correspond to the three different values proposed by Bertrand. It is crucial however to observe that these three averages are now the solutions of three very different problems, referring to three very different entities. Therefore, and exactly the other way around, it would be paradoxical if the numerical value of these three averages would coincide. 

Of course, it is always possible to consider a more abstract variant of BQ, expressing a deeper level of randomness (i.e., of potentiality), which is the following:\\

\noindent {\bf  Modified Bertrand's Question (BQ4)}: We choose \emph{at random} one of the following three processes, then perform it: (1) we draw \emph{at random} a chord onto a circle; (2) we draw \emph{at random} two points onto a circle and draw a chord passing through them; (3) we draw \emph{at random} a point inside a circle and a chord having such point as its middle point. What is the probability that the obtained chord is longer than the side of the inscribed equilateral triangle?
\\

This BQ4 version expresses a deeper level of randomness with respect to the questions BQ, BQ2 and BQ3. Indeed, in BQ, BQ2 and BQ3, only two processes of randomization are involved. The first one corresponds to the selection of a ``$\rho$-way of tossing'' the entity under consideration, and the second one, subsequent to it, corresponds to the selection of a given interaction, which then deterministically produces the final chord. Two averages are therefore involved: a \emph{uniform average} over all possible $\rho$, and a subsequent \emph{weighted average}, with the weights which are determined by the actualized $\rho$. 

On the other hand, in BQ4 there is an additional layer of ``actualization of potential elements of reality,'' which corresponds to the non-deterministic selection of an entity from a set of three different potential entities. Here we can either directly apply the principle of indifference to the three possible outcome-entities, or consider an average over all possible ways of choosing an entity out of a set of three entities, which will still produce the same response. Thus, the answer to BQ4 is:
\begin{equation}
{\cal P}(L>\sqrt{3}) =  {1\over 3}\left[{1\over 2} + {1\over 3} + {1\over 4}\right]= {13\over 36}.
\label{solution4}
\end{equation}

We would like to conclude the present work by offering a different perspective on Bertrand's problem, related to the possibility of generally understanding probability theory as a theory dealing with the measurement of properties of physical entities, operationally defined by means of specific observational protocols. 

For instance, we can consider  EQ as the operational definition of a property of the stick, which we may  call the \emph{equilaterality} property, considering that when the property is actual the chord generated by the stick is contained into an equilateral triangle.  In other terms, we can consider EQ as the experimental \emph{test} for such \emph{equilaterality} property, in the sense that, if, by tossing a stick, we find a chord of length $L>\sqrt{3}$, we will say that the observation of the equilaterality property has been successful, and unsuccessful otherwise. 

Now, since we cannot predict with certainty what will be the outcome of the above measurement, \emph{equilaterality} will be in general only a potential property of the stick, which can only be actualized (that is, created) during its measurement process, with probability ${1\over 2}$, as it is easy to confirm in concrete experiments.~\cite{Jaynes, Porto}

This situation is very similar to  the measurement of properties of microscopic entities, like for instance the spin of an electron. Indeed, independently of our knowledge of the state of an electron, we are generally only in a position to predict the outcomes in statistical terms. A possible reason for this is that we lack knowledge about the interaction which is each time actualized between the measuring apparatus and the measured entity, so that  a quantum measurement would be the expression not of a single interaction, but of an entire collection of potential measurement interactions.

This explanation of  quantum indeterminism in terms of ``hidden'' measurement interactions, was firstly proposed by one of us, several years ago, as a possible solution to the \emph{measurement problem}, as it can be shown that such explanation is fully compatible with the predictions of the quantum mechanical Born rule.~\cite{Aerts1986} In the development of this ``hidden-measurement approach,'' it was clear from the beginning that the existence of hidden variables associated with the measuring interactions was able to generate additional probability models, not necessarily describable by the Born rule. Consequently, the idea emerged that a quantum measurement could  also be the expression of a much deeper level of potentiality (and therefore of randomness), describing not only a process of actualization of potential interactions, but also of actualization of entire ways of choosing these interactions. 

This idea was presented in Refs.~\onlinecite{Aerts1998, Aerts1999b}, and measurements expressing such a deep level of potentiality were called \emph{universal measurements}. The conjecture was that quantum measurements precisely constitute an example of universal measurements, i.e., that the quantum statistics delivered by the Born rule would be the result of a uniform average over all possible ``ways of selecting an interaction,'' which in the present work we have called a \emph{universal average}. At that time, however, it was not clear how to handle, mathematically speaking, such a huge average, and this for difficulties precisely related to the Bertrand paradox. 

Only very recently it became clear that a discretization process ``\`a la Wiener'' was the right way of addressing the problem, thus allowing to  show that, when the state space is Hilbertian, universal measurements produce exactly  the same statistics as that delivered by the Born rule.~\cite{AertsSassoliOne, AertsSassoli,AertsSassoli-Bloch} As a corollary of this result, we obtained what we think is also a convincing solution to Bertrand's paradox, which we have tried to present, explain and motivate in this article.

This intimate connection between fundamental problems of probability theory, like Bertrand's paradox, and of quantum mechanics, like the measurement problem, is certainly not coincidental. Both approaches are  aimed at the description of systems subjected  to  specific experimental actions, according to  protocols which incorporate the presence of fluctuations in the experimental context, so that the outcome of these actions cannot be predicted in advance with certainty, not even in principle. Also the simple act of tossing a stick, or a die,~\cite{Sassoli} when considered from that broader perspective, becomes representative of a quantum processes, so that we could say that the founding fathers of probability theory, without knowing it, were actually quantum physicists \emph{ante litteram}!

\end{document}